# Architectural Consistency Checking in Plugin-Based Software Systems

Timo Greifenberg
Software Engineering
RWTH Aachen University
http://www.se-rwth.de
greifenberg@se-rwth.de

Klaus Müller
Software Engineering
RWTH Aachen University
http://www.se-rwth.de
mueller@se-rwth.de

Bernhard Rumpe
Software Engineering
RWTH Aachen University
http://www.se-rwth.de
rumpe@se-rwth.de

## ABSTRACT

Manually ensuring that the implementation of a software system is consistent with the software architecture is a laborious and error-prone task. Thus, a variety of approaches towards automated consistency checking have been developed to counteract architecture erosion. However, these approaches lack means to define and check architectural restrictions concerning plugin dependencies, which is required for plugin-based software systems.

In this paper, we propose a domain-specific language called Dependency Constraint Language (DepCoL) to facilitate the definition of constraints concerning plugin dependencies. Using DepCoL, it is possible to define constraints affecting groups of plugins, reducing the required specification effort, to formulate constraints for specific plugins only and to refine constraints. Moreover, we provide an Eclipse plugin, which checks whether the software system under development is consistent with the modeled constraints. This enables a seamless integration into the development process to effortless check consistency during development of the software system. In this way, developers are informed about dependency violations immediately and this supports developers in counteracting architecture erosion.

## Categories and Subject Descriptors

D.2 [**Software Engineering**]: Software Architectures

## 1. INTRODUCTION

The software architecture of a software system defines the structuring of a software system into components and how these components are expected to interact [12, 7]. Furthermore, the software architecture defines design decisions that are crucial for understanding a software system [15]. However, if the architecture of a software system is documented at all, the architectural descriptions are seldom maintained [9]. This stems from the fact that manually ensuring that the software architecture stays consistent with the implementation of a software system is a laborious and error-prone task. This deviation of architecture and implementation, known as architecture erosion [12], negatively impacts software quality, maintainability and evolvability [9, 16, 8].

To cope with this problem, a variety of approaches have been developed, allowing to check this consistency automatically [9, 11, 2]. These approaches do, for instance, provide means to define architectural rules that have to be fulfilled or that must not be violated by the implementation. Other approaches are based on a definition of the expected architecture and an extraction of the existing architecture [10, 3]. If a deviation is detected, developers are informed about these. However, these approaches mainly focus on (fine-grained) dependencies between source code entities such as classes and do not allow to easily define more abstract constraints concerning dependencies between plugins. Expressing comparable constraints with the existing approaches is possible to a certain extent, but more laborious and less convenient for a system architect having to define constraints for plugin dependencies. In particular, this paper assumes that the plugin-based software system whose consistency is checked was developed with Eclipse [6] and, thus, consists of Eclipse plugins and Eclipse features. Despite this, our approach is transferable to plugin-based systems developed with another tooling infrastructure as long as dependencies between plugins and groups of plugins can be identified.

This paper proposes a domain-specific language (DSL), called Dependency Constraint Language (DepCoL), for defining constraints concerning plugin and feature dependencies in a plugin-based software system. By providing an easy to use and convenient DSL for software architects of plugin-based software systems, we assume that there is a higher chance that the approach is accepted and permanently integrated into the development process. Using DepCoL, it is possible to define constraints for single elements as well as groups of elements, to reduce the specification effort. To process constraints modeled in DepCoL, an Eclipse plugin was developed which can be integrated into the development environment of the plugin-based software system. This plugin is capable to check the consistency between the implementation of the system under development and the modeled constraints automatically. These checks can be performed accompanying the development of the software system. If a developer introduces a dependency which is forbidden according to the constraint specifications, this violation is di-rectly reported to the developer so that the violation can be fixed instantly.

The paper is structured as follows: in Section 2, we elaborate on the background of our work in which we introduce important terms and the application context of DepCoL. After that, we give an overview on our approach in Section 3. DepCoL is then presented in detail in Section 4. Section 5 then describes related work before we summarize the paper in Section 6.

## 2. BACKGROUND

To avoid misunderstandings, relevant term definitions are introduced first. As DepCoL was developed in a cooperation project with an industrial partner, the application context of this work is briefly described afterwards.

### 2.1 Term definitions

This paper focuses on plugin-based software systems developed in Eclipse. Due to this, it is concerned about dependencies between Eclipse plugins and Eclipse features. In the further course of this paper Eclipse plugin will be abbreviated as plugin and Eclipse feature as feature.

The smallest component of a plugin-based system that is considered in this work is a plugin. In simple terms, a plugin is a collection of classes and libraries required by these classes. Multiple plugins can be grouped in a feature. In Eclipse, a feature can also contain other features but to simplify the description, it is assumed in the following that features merely contain plugins. In addition, this approach uses the term feature group to refer to a logical group of features. The user is free to define which features should be contained in such a feature group. Analogous to that, a plugin group represents a logical group of plugins.

Based on these definitions, both features and plugin groups are composed of plugins. The difference between both terms is that a feature is an essential part of the plugin-based software system itself, whereas a plugin group is only defined in the dependency model to group selected plugins. To be able to distinguish between these two kinds of plugin groups on the level of the dependency model, different terms are used in this work.

Each plugin might depend on a multitude of other plugins in order to work properly. This means that the plugin can only be executed, if these plugins are also available. Consequently, there are (indirect) dependencies between plugins and features, between features and features and so forth. The term dependency is concretized in the following.

A plugin "m" depends on a plugin "n" if "n" is listed as a required bundle in the Eclipse manifest file of "m". All other dependencies referred to in this work are based on this dependency definition. Before this is explained in more detail, please note that, in the end, a plugin, a feature, a plugin group and a feature group can all be represented as a group of plugins. For a feature and a plugin group this is obvious. A single plugin can be represented by a plugin group containing only the single plugin. For a feature group, the resulting group of plugins can be derived by collecting the plugins of all features contained in the feature group.

Let $X$ and $Y$ represent a plugin, a feature, a plugin group or a feature group and let $PG$ be a function which returns the group of plugins for the passed argument. Then, $X$ depends on $Y$ if there is a plugin $m \in PG(X)$ and a plugin $n \in PG(Y)$ such that plugin "m" depends on plugin "n".

### 2.2 Application context

The dependency constraint language DepCoL was developed in a cooperation project with an industrial partner. This cooperation project concerned complex Eclipse rich client applications which are together composed of roughly 800 plugins and 100 features and which share several features and plugins.

In order to support the evolution of these plugin-based software systems, the goal of the cooperation project was to develop a DSL which is capable to define constraints concerning the dependencies between features and plugins of these software systems. One special requirement was that it should be possible to define constraints on differing levels of granularity. Not only should the language be capable to express restrictions for groups of elements but also specific restrictions for specific elements. Another requirement was that developers should be informed about violations against defined constraints as soon as possible. Section 4 introduces the resulting dependency constraint language DepCoL.

## 3. OVERVIEW

The key idea that underlies the presented approach is that there is an explicit dependency model which contains constraints concerning dependencies between plugins and features of a plugin-based software system. The consistency between the implementation of a software system and the dependency model can be checked automatically. Figure 1 gives an overview of the basic workflow which is established through the presented approach.

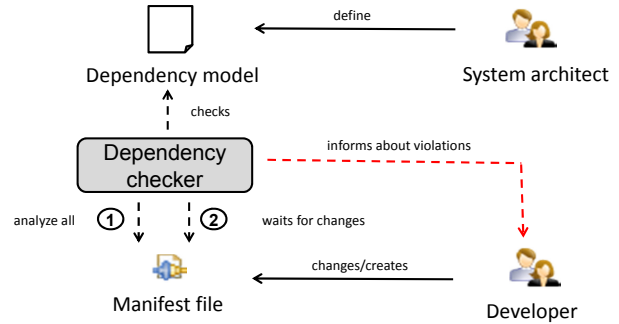

**Figure 1: Overview of architecture consistency checking process**

At the beginning, the system architects have to define a dependency model using DepCoL. This dependency model encapsulates the architectural rules that must not be violated by the analyzed plugin-based software system.

The dependency checker that is responsible for automatically checking the architectural consistency needs to be integrated into the development environment in which the plugin-based software system to be checked is developed. Then, the dependency checker can automatically check for violations of that software system against the given dependency model. For this purpose, at first the existing dependencies between the plugins of the plugin-based software system are extracted. As this work focuses on plugin-based systems which are developed with Eclipse, the extraction of the existing dependencies is carried out by analyzing Eclipse manifest files. The dependency checker either extracts all dependencies from the Eclipse manifest files belonging to the corresponding plugin-based software system or it is triggered to extract the dependencies of one specific plugin (and the associated feature) only. The latter case is performed in case the dependency checker was informed that the dependencies of one particular plugin have been changed. In this situation, the dependency checker only needs to check whether the changed plugin violates the dependency model and, hence, only the dependencies of this changed plugin need to be obtained.

Figure 2: Eclipse view showing results of dependency checker

No matter in which way the invocation of the dependency checker is initiated, the developers are informed immediately about violations against the dependency model through an Eclipse view as shown in Figure 2. In this view, the violation messages are grouped by the severity of the violation. If a dependency to a feature is detected as forbidden or tolerated, the view also indicates the dependencies to which plugins of that feature cause the violation. In addition to that, the manifest files of the plugins violating the dependency model are annotated with an error marker. This instant feedback allows for directly resolving the undesired dependencies and thus directly counteracts architecture erosion.

## 4. THE DEPENDENCY CONSTRAINT LANGUAGE DEPCOL

In this section, the dependency constraint language DepCoL is presented. DepCoL provides two kinds of language constructs in order to define a dependency model: those allowing to define structures that refer to plugins or features and those that make it possible to formulate constraints concerning dependencies between plugins, features and the previously defined structures. An example for the definition of a structure referring to features is a feature group. In its definition, the user has to state which features belong to the feature group. Listing 2 on the next page shows a concrete example for feature group definitions. As soon as the user introduced a structure such as a feature group, concrete constraints can be declared for the elements contained in the structure. For instance, it could be formulated that all features from one feature group must not depend on all features from an other feature group.

In the following, at first the different language constructs of DepCoL are presented in more detail. A self-contained but fictive example for a dependency model can be found at [5]. After that, the semantics of a dependency model is briefly described. Finally, the implementation of the dependency checker is sketched.

```
declare featurebase {
  f1;
  f2;
  f3;
}
```

Listing 1: Feature base example

### 4.1 Definition of structures referring to features and plugins

#### 4.1.1 Feature and plugin base

A dependency model usually refers to a variety of feature and plugin names. To avoid that feature or plugin names are written in different ways in the dependency model, the concept of a feature base and plugin base was introduced.

A feature base contains the names of all features which can be directly referenced in the dependency model. A context condition ensures that only feature names which are listed in the feature base can be named in the dependency model. Listing 1 shows an example for a feature base specification, through which it is declared that the features "f1", "f2" and "f3" can be used in the dependency model.

Analogous to that, all plugin names which can be referenced have to be declared in a plugin base. Once again, this is ensured by a context condition. In order to support the developers, the provided Eclipse tooling can generate feature and plugin bases which contain all feature and plugin names from a particular workspace.

#### 4.1.2 Feature and plugin group

To simplify the specification of constraints for multiple features and plugins, DepCoL supports feature and plugin groups. A feature group can contain a list of feature names that are defined in the feature base. In addition to that, a feature group can contain regular expressions for feature names. Within a regular expression, it is possible to make use of the wildcard "*" which represents an arbitrary string. Finally, a feature group can contain other feature groups. In summary, this means that a feature "f" is contained in a feature group "fg" if one of the following conditions is fulfilled:

- feature "f" is explicitly listed in the feature group "fg".

- feature "f" matches a regular expression contained in the feature group "fg".

- feature "f" is contained in a feature group "subfg" that is contained in "fg".

```
1  declare featuregroup fgListFeatures {
2    f1;
3    f2;
4  }
5  declare featuregroup fgRegExp {
6    fs.ui.*;
7    fs.core.*;
8  }
9  declare featuregroup fgCombined {
10    f3;
11    fs.ext.*;
12    featuregroup fgListFeatures;
13    featuregroup fgRegExp;
14  }
```

**Listing 2: Different variants for feature group specifications**

```
1  plugin p1 {
2    [critical] forbid dependency to feature f2;
3    [warning] forbid dependency to featuregroup
4      fg2;
5  }
6  featuregroup fgListFeatures {
7    tolerate dependency to featuregroup
8      fgRegExp;
9  }
```

**Listing 3: Examples for forbidding/tolerating dependencies**

Please note that, as each feature name listed explicitly in a feature group can be regarded as a simple regular expression and a feature group can contain further regular expressions, each feature group can in the end be represented as one regular expression which potentially consists of multiple alternatives and which represents a subset of features. Analogous to that, each plugin group can be represented as one regular expression.

Examples for feature group specifications are given in Listing 2. In this example, the feature group "fgListFeatures" only contains the features "f1" and "f2". The feature group "fgRegExp" refers to all features with names that start with "fs.ui." or with "fs.core.". Finally, feature group "fgCombined" illustrates how different notations can be combined. The feature group contains the features from the previously described feature groups and in addition, feature "f3" as well as all features with names that start with "fs.ext.".

Analogous to the feature group specification, a plugin group can comprise multiple explicitly stated plugin names, regular expressions for plugins and other plugin groups. Feature and plugin groups do not have to be disjunct, i.e., a feature can be contained in multiple feature groups and a plugin can be contained in multiple plugin groups.

Amongst others, feature and plugin groups can be used to represent layered software architectures [1]. For each layer, a feature or plugin group can be created which comprises the features or plugins which are contained in the corresponding layer. Moreover, constraints can be defined for each feature or plugin group to restrict the dependencies between the layers. The definition of constraints will be explained in the next subsection.

## 4.2 Definition of dependency constraints

Constraints can be formulated for all features included in the feature base and all plugins named in the plugin base. Furthermore, constraints can be stated for all feature groups and plugin groups. For each of these elements, it can be defined whether dependencies to specific features, plugins, feature groups or plugin groups are forbidden, tolerated or allowed. By default, dependencies between all elements are allowed. Due to this, it is necessary to define forbidden and tolerated dependencies explicitly. Nevertheless, it can be required to explicitly allow a dependency, which is explained in more detail later in this subsection. This default behavior can also be adapted if desired.

Tolerated dependencies can be used to indicate that a dependency is not desired but also not (yet) forbidden, temporarily allowed but potentially forbidden in future or in case a dependency modeler is unsure about whether the dependency should be allowed or forbidden. By default, the user receives a warning message if a dependency is detected by the dependency checker that is defined as tolerated. In contrast to this, an error message is reported if a forbidden dependency is detected. For forbidden dependencies, it is possible to denote a severity on how critical a violation against the specification is. This severity embodies how important it is to fix a violation. Possible severity types are: critical, error and warning. In the reporting of the violation messages, the messages are differentiated according to their severity type, as shown in Figure 2.

Please note that, for the dependency checker, forbidden dependencies with severity warning and tolerated dependencies are treated equally, as both result in warning messages. However, for a dependency modeler, there is a difference between both. A forbidden dependency with severity warning is really forbidden, but a violation is not critical and it is sufficient to report a warning. A tolerated dependency instead usually expresses, that the dependency is not desired and that it needs to be clarified how to handle the dependency.

Listing 3 shows examples for concrete dependency constraints. These indicate that dependencies from plugin "p1" to feature "f2" and to all features contained in feature group "fg2" are forbidden. Violations against the first constraint are regarded as critical, whereas violations against the second constraint are regarded as warnings. Moreover, Listing 3 indicates that dependencies from features contained in feature group "fgListFeatures" to features contained in feature group "fgRegExp" are only tolerated.

An important property of DepCoL is that constraints can refine previously defined constraints for specific features, plugins, feature groups and plugin groups. In this way, it can, e.g., be expressed that all plugins in a plugin group must not depend on specific plugins and after that, it can be refined that selected plugins are allowed to depend on specific other plugins. The refinement concept is explained in more detail in the next subsections.

## 4.3 Semantics of a dependency model

As explained in Subsection 2.1, a plugin, a feature, a plugin group and a feature group can all be represented as a group of plugins in the end. Let $X$ and $Y$ refer to a plugin, a feature, a plugin group or a feature group and let the function $PG$ return the group of plugins represented by the passed argument. Then, a constraint between $X$ and $Y$ can be represented as a set of pairs of plugins $(x_i, y_i)$ for all $x_i \in PG(X)$ and $y_i \in PG(Y)$. Each constraint in the dependency model can be described by such pairs of plugins.

Moreover, each plugin pair is assigned a property, that indicates whether the plugin pair represents a dependency which is regarded as critical, an error, a warning or allowed. By default, all dependencies are allowed. Thus, the allowed property is initially assigned to all possible plugin pairs. For each constraint, this property is changed according to the constraint type. For instance, for a forbidden dependency with severity error, the according plugin pairs are assigned the property error.

Let $c_1$ and $c_2$ be two constraints in the dependency model and let $PP$ denote the function which returns the plugin pairs for each constraint. A constraint $c_2$ refines another constraint $c_1$ if and only if $PP(c_1) \cap PP(c_2) \neq \emptyset$ and if $c_1$ is defined before $c_2$. The intersection between $PP(c_1)$ and $PP(c_2)$ contains exactly the refining plugin pairs. If one constraint refines another constraint, the refining constraint determines the property associated to the refining plugin pairs. If, e.g., a constraint $c_1$ describes pairs of plugins between which dependencies are regarded as errors and a constraint $c_2$ describes pairs of plugins between which dependencies are regarded as critical and $c_2$ refines $c_1$, then all refining pairs of plugins are assigned the property critical.

To sum it up, a dependency model describes a mapping from pairs of plugins to a property which indicates whether a dependency between the plugins is regarded as allowed, critical, as an error or as a warning.

## 4.4 Dependency constraint checker implementation

In this section, the implementation of the dependency checker is outlined and illustrated by a concrete example. It is assumed that it is checked whether the dependencies of a particular plugin and the feature containing that plugin violate a given dependency model. The case that the dependencies of all plugins are validated is handled by performing these steps for each plugin.

### 4.4.1 Implementation

As a first step in the dependency checking phase, the dependency checker extracts the existing dependencies between the plugins that are part of the analyzed software system. This extraction is only performed if the plugin dependencies have not been extracted before or the plugin dependencies have changed since the last extraction.

The most important aspect that has to be considered in the dependency checker is that constraints can be refined by constraints that are defined later in the dependency model. To cope with this, the dependency model is processed from end to beginning and each structure in which constraints are defined is traversed from end to beginning too. In this traversal of the dependency model, a dependency is only considered, if it has not been addressed by a previously processed constraint, as will be explained in more detail subsequently. Moreover, only those constraints are treated that concern the validated plugin or the feature containing the validated plugin.

While processing the relevant constraints, the dependency checker builds up relations for the pairs of plugins between which dependencies are allowed or between which dependencies are regarded as critical, as errors or as warnings according to the dependency model. These four relations are named $R_{allowed}$, $R_{critical}$, $R_{error}$ and $R_{warning}$ in the following and realize the mapping from the pairs of plugins which are created for each constraint to the dependency type property, as described in the previous subsection. To simplify the description, the relation $R$ refers to the union of all four relation variants subsequently.

One speciality of the implementation is that for each constraint only pairs $(x_i, y_i)$ are considered in which $y_i$ is a plugin required by $x_i$. Consequently, $R$ only contains information concerning the plugins which are required by the analyzed plugin. A further important aspect of the implementation is that $R$ only stores pairs of plugins which were created for the explicitly defined constraints. If there is no constraint concerning a dependency between two plugins $x_i$ and $y_i$, the pair $(x_i, y_i)$ is not contained in $R$ and it is implicitly treated as allowed.

Due to the fact that the dependency model is traversed from end to beginning and constraints can refine other constraints, a pair $(x_i, y_i)$ is only added to $R$, if $R$ does not already contain the pair $(x_i, y_i)$. Otherwise a constraint has been processed before, which has refined the currently processed constraint. As a consequence of this, all $R$ relations are disjoint by construction.

At the end, the dependency checker traverses the relations $R_{critical}$, $R_{error}$ and $R_{warning}$ and creates violation messages for the contained entries.

Besides managing the pairs of plugins in $R$, it is stored which concrete constraint led to the addition of the corresponding plugin pair. One use case for this is that it facilitates the creation of violation messages. The relation $R_{warning}$ contains pairs of plugins between which dependencies are regarded as warnings. However, on this level, it cannot be differentiated between pairs which were created due to a tolerated or a forbidden dependency with severity warning. By managing the constraints which led to the addition of the pair, it is possible to differentiate both cases. Consequently, the resulting violation message can specify whether the dependency is forbidden or tolerated. Moreover, it helps to construct better violation messages. If, e.g., the dependency model contains a constraint which expresses that a feature is forbidden to depend on another feature, a violation message can be created which expresses exactly this, instead of only reporting that the dependency between two plugins is forbidden. The resulting violation messages, thus, reflect the specifications in the dependency model.

One further use case for managing which constraint led to the addition of which plugin pair is that this allows for informing the users about which constraint refines which other constraint. Each time, a pair $(p_i, p_j)$ should be added to $R$, but $R$ already contains that pair, it can be logged that the constraint which is mapped to that stored pair has refined the currently processed constraint. By reporting this log to the users, the dependency modeler can verify if the refining is done on purpose or not. The motivation for this is that especially in large dependency models a dependency modeler might not always be aware of the fact that a constraint refines previous constraints.

### 4.4.2 Example

The processing of the dependency model given in Listing 4 is sketched in the following. In this example, it is assumed that the plugin base, denoted in the following by $PB$, contains the plugins "p1", "p2", "p3", "p4.ui", "p5.ui" and "p6.i18n". Moreover, it is assumed that plugin group "pg1" contains at least the plugin "p1", that plugin group

```
1  plugingroup pg1 {
2    forbid dependency to plugingroup ALL;
3    tolerate dependency to plugingroup pgUi;
4  }
5  plugin p1 {
6    allow dependency to plugin p4.ui;
7  }
```

**Listing 4: Example for plugin refining plugin group constraints**

"pgUi" contains all plugins with names ending with ".ui" and that plugin group "ALL" contains all plugins. Finally, it is assumed that plugin "p1" is validated and that "p1" requires the plugins "p4.ui", "p5.ui" and "p6.i18n".

As plugin "p1" is validated, only constraints for this plugin are processed. Due to the constraint in line 6, the pair $(p1, p4.ui)$ is added to $R_{allowed}$. In the next step, the constraints for plugin group "pg1" would be processed, starting with the constraint in line 3. Based on this constraint, it would be checked whether the pairs $(p1, p4.ui)$ and $(p1, p5.ui)$ can be added to $R_{warning}$. As $R_{allowed}$ already contains $(p1, p4.ui)$, only $(p1, p5.ui)$ is added to $R_{warning}$. Finally, the pairs $(p1, p_x)$ with $p_x \in PB$ are added to $R_{error}$, provided that $p_x$ represents a required plugin and $R$ does not already contain the corresponding pair. Therefore, only $(p1, p6.i18n)$ is added. As a result, a warning message would be reported for the tolerated dependency from "p1" to "p5.ui" and an error message would be reported for the dependency between "p1" and "p6.i18n".

## 5. RELATED WORK

In this section, we discuss related work in the context of static architecture consistency checking. The most important difference between our work and the existing approaches is that our work focuses on plugin-based software systems. In contrast to this, most existing approaches address dependencies on the source code level.

In [14, 15], dependencies within a software system are restricted using the static, declarative language DCL. DCL provides means to define constraints between modules, which represent sets of classes. In order to specify which classes belong to a module, the classes can either be listed by denoting their fully qualified names or by using regular expressions which can match multiple classes. Based on these models, several constraints can be defined, e.g., that only classes from module A can access classes from module B. The definition of modules in [14, 15] is comparable to the definition of feature and plugin groups.

In [10], the software reflexion model technique is introduced. In this technique, developers have to define a high level model which contains entities and relations between these entities. Moreover, developers have to extract a source model such as a call graph from the source code and at the end, a mapping between the high-level model and the source model has to be performed. In the mapping phase, entities from the high-level model have to be assigned to elements of the source model, typically using regular expressions. Based on these models and the mapping, a software reflexion model is computed to determine where the high-level model does (not) comply with the source model. In contrast to reflexion models, in our work developers do not have to provide a high-level model and a mapping to the source model. Instead, constraints are defined that express allowed, tolerated or forbidden dependencies. Furthermore, it is easily possible to refine other constraints.

In [13], dependencies are extracted from the source code by means of static code analysis. The resulting dependencies are shown in a dependency structure matrix (DSM). To define what kinds of dependencies in the DSM are allowed or forbidden, design rules can be defined. These are applied to a DSM to identify which dependencies violate the intended software architecture. By default, design rules are inherited, i.e., if restrictions are defined for a certain subsystem, all elements of that subsystem inherit these restrictions. Moreover, design rules can be refined for specific elements of the DSM. The idea of refining constraints in our work was motivated by the possibility to refine design rules. Moreover, constraints are also inherited, e.g, if constraints are defined for a feature group, by default, all plugins of the contained features inherit these constraints. However, developers can also express tolerated dependencies in DCL.

One further strategy to cope with architecture erosion is to use query languages [18]. Queries can be executed on the source code or other relevant artifacts to, e.g., identify all classes that depend on particular other classes or packages [17]. Thus, to assess the architectural consistency, queries need to be written that identify forbidden dependencies.

For example, [4] applies a rule-based approach for architecture compliance checking. This approach is based on formal notations to represent the intended software architecture, the design or implementation of the system and mappings between both. Each of these parts is represented as a logical knowledge base which contains facts that represent the elements of the descriptions. By executing the architectural rules on the union of the knowledge bases, the architectural compliance is checked. In particular, constraint rules can not only report a violation against a rule but also identify the violating elements. In [8], this idea is refined. Architectural rules are defined as logical formulas on a common extensible ontology and models are mapped to instances of this ontology. To represent models in this form, transformations have to be defined that state how model elements can be represented. This makes it possible to use models to define architecture rules. Thus, knowledge about logical programming is mainly required by the developers defining the transformations. In contrast to these works, our work does not require any knowledge about logic programming at any level, on the other hand, it is less flexible. However, this flexibility is not required in our use case as we focus on plugin-based systems on purpose. In this way, we can focus on aspects known by developers and architects who are familiar with Eclipse.

Further approaches and tools concerning static architecture compliance checking are presented in [11, 9]. A detailed survey on architecture erosion and approaches to minimize, prevent and repair architecture erosion is given in [2].

As dependencies between plugins are usually rooted in the fact that classes of the plugins somehow depend on each other, an alternative to using our implementation of the dependency checker would be to transform a dependency model into a model that is suitable for a class-based approach. The advantage of this would be that the resulting model could be checked with the existing class-based tooling. One problem in this scenario would be that an Eclipse manifest file can contain unused dependencies, which can-not be traced back to class dependencies. Hence, a class-based approach cannot detect these dependencies. To circumvent this problem, it must be ensured that no manifest file contains unused dependencies. A further potential problem would be that multiple plugins can contain a class with the same name - even though this would be bad practice. A class-based approach could not differentiate between these classes. Moreover, it would be required to transform the dependency model to the class-based model after every manifest file change. These reasons and the requirements for supporting tolerated dependencies, constraint refinements and the integration in Eclipse resulted in the decision to implement our own dependency checker tailored to our DSL.

## 6. CONCLUSION

In this paper, we have proposed the DSL DepCoL to define constraints concerning plugin and feature dependencies in a plugin-based software system. With DepCoL, constraints can be specified on different levels of granularity, e.g., for feature and plugin groups but also for specific features and plugins. One speciality of DepCoL is that it supports the refining of constraints. This facilitates defining some general constraints first which can then be refined for specific elements, if required.

By integrating a dependency checker into the development environment in which the plugin-based software system is implemented, developers can be informed immediately in case the changes they performed violate the previously defined constraints. In this way, DepCoL and the according tooling counteract architectural erosion in plugin-based software systems. This approach is mainly feasible for complex plugin-based software systems with a multitude of plugins. In such systems it is not possible to keep track of the dependencies between plugins and features manually.

For future work, we plan to conduct a long-term case study in the industrial cooperation project. In the course of this, we particularly plan to investigate whether the refinement interpretation we are using so far is suited for complex plugin-based software systems or whether other information should be used, e.g., the hierarchy of elements. Moreover, we plan to analyze whether it would be advantageous to consider transitive dependencies which are not considered so far.